\documentclass[pra,twocolumn]{revtex4b5} 
\begin{document}

\bibliographystyle{unsrt}

\title{Probabilistic Quantum Logic Operations Using Polarizing Beam Splitters} 
\author{T.B. Pittman,  B.C. Jacobs, and J.D. Franson}
\affiliation{Applied Physics Laboratory, The Johns Hopkins University, Laurel,
MD 20723} 
\date{July 18, 2001}

\begin{abstract}
It has previously been shown that probabilistic quantum logic operations can be performed
 using linear optical elements, additional photons (ancilla), and post-selection based on
the output of single-photon detectors.  Here we describe the operation of several  
quantum logic operations of an elementary nature, including a quantum parity check and 
a quantum encoder, and we show how they can be combined to
implement a controlled-NOT ({\scriptsize CNOT}) gate.  All of these gates can be
constructed using polarizing beam splitters that completely transmit one state of polarization 
and totally reflect the orthogonal state of polarization, which allows a simple
explanation  of each operation.  We also describe a polarizing beam splitter  
implementation of a {\scriptsize CNOT} gate that is closely analogous to
 the quantum teleportation
technique previously suggested by Gottesman and Chuang [{\em Nature {\bf 402},
390 (1999)}].  Finally,
our approach 	     
has the interesting feature that it makes practical use of a quantum-eraser
technique.
\end{abstract}

\maketitle

\section{Introduction}

 An optical approach to quantum information processing would have
several advantages, including the ability to transport qubits from one location
to another using optical fibers or wave guides.  The main difficulty with any
optical approach is that nonlinear interactions between individual photons are
required in order to implement quantum logic gates that operate with 100\%
efficiency.  It was recently shown in two pioneering papers
\cite{knill01,koashi01},
 however, that probabilistic quantum logic gates 
 can be implemented using linear optical
elements, additional photons (ancilla), and post-selection based on the output of
single-photon detectors.  Probabilistic logic gates of this kind will give the
desired result with certainty when a specific output from the detectors is
obtained, but that will only occur for some fraction of the events, typically
with a probability of $\frac{1}{16}$ to $\frac{1}{4}$.  Somewhat remarkably,
Knill, LaFlamme, and Milburn (KLM) \cite{knill01}
 have shown that the probability of
success of  a given operation can
approach unity in the limit of large numbers of ancilla and detectors
\cite{kwiatlinear}.

In this
paper, we describe a variety of quantum logic gates that can be constructed using
polarizing beam splitters that completely transmit one state of polarization and
totally reflect the orthogonal state of polarization.  One advantage of this
approach is that it enables us to give simple explanations for the operation of all of the
relevant quantum logic gates. 
In addition, polarization-encoded qubits are typically less sensitive to phase
drifts than are interferometric implementations.
We describe several kinds of probabilistic quantum
logic gates, including a quantum parity check and a quantum encoder, along with
two different implementations of the controlled-NOT ({\scriptsize CNOT}) gate.

The original two approaches \cite{knill01,koashi01}
 for implementing probabilistic quantum
logic gates
using linear elements are closely related to an earlier paper by Gottesman and
Chuang (GC) \cite{gottesman99}, who showed that a {\scriptsize CNOT} gate could
be implemented using a modified form of quantum teleportation \cite{bennett93} and a
four-qubit entangled state as a resource
(ancilla).  The Bell-state measurements \cite{braunstein92} required for
teleportation are inherently nonlinear \cite{vaidman99,lutkenhouse99,kim01}, 
but a partial set of Bell-measurments can be performed 
probabilistically using linear optical elements
\cite{michler96,mattle96,bouwmeester97}. 
 Combining these two ideas leads to a {\scriptsize CNOT} gate that we describe in
Section II using polarizing beam splitters.  As we already mentioned, this
approach provides a straightforward way to understand the operation of the logic
gate as well as some potential experimental advantages in its operation.  This
{\scriptsize CNOT} gate is similar in some respects
 to those previously described by 
 KLM \cite{knill01} as well as by Koashi, Yamamoto, and Imoto 
(KYI) \cite{koashi01}. 

In Section III, we introduce several simple
quantum logic gates that combine two photons on a single polarizing beam splitter
to perform a variety of elementary functions, including a 
destructive-{\scriptsize CNOT} 
gate in addition to the quantum parity check and quantum encoder
mentioned above.  These elementary operations are then combined to give an alternative
implementation of a complete probabalistic {\scriptsize CNOT} gate that
does not rely on teleportation in any obvious way. 

Our
approach makes use of a technique that can be viewed as being equivalent to a
quantum eraser \cite{scully82,kwiat92,herzog95}, as is described in the
Appendix.  We summarize our results and consider the prospects for  
future applications in Section IV.

\section{{\small CNOT} using Four-Photon Entangled States}

As we mentioned earlier, Gottesman and Chuang showed
in a pioneering paper  \cite{gottesman99} that a {\scriptsize CNOT}
 operation could be performed using a modified
form of quantum teleportation.  Although the required
Bell-state measurements are inherently nonlinear
\cite{vaidman99,lutkenhouse99,kim01},  a partial or incomplete set 
of Bell state measurements
can be implemented using linear optical elements and post-selection, as has been
demonstrated experimentally by Zeilinger's group
\cite{michler96,mattle96,bouwmeester97}.  
A combination of these two techniques might be expected to allow the implementation
 of a probabilistic
{\scriptsize CNOT}, which formed part of the motivation for earlier work in this area
\cite{knill01,koashi01}. 
Here we describe an implementation of a probabilistic {\scriptsize CNOT} along these lines
using polarizing beam splitters and a quantum erasure technique.  This approach
requires ancilla in the form of a four-photon entangled state $|\chi\rangle$ 
that can be created in various ways \cite{knill01,koashi01,gottesman99}, as will be
discussed further in Section IV below.  The implementation given here is 
similar in spirit to those of Refs.  \cite{knill01}
and  \cite{koashi01}, but our approach allows a more straightforward
understanding of the operation of the {\scriptsize CNOT} gate and
 may have some practical advantages as well.

\begin{figure}[b]
\vspace*{-.25in}
\begin{center}
\begin{picture}(200,170)
\thinlines
\put(10,60){\line(1,0){60}}
\put(40,30){\line(0,1){60}}
\thicklines
\put(40,60){\vector(1,0){30}}
\put(40,60){\vector(0,1){30}}
\put(40,60){\vector(1,1){20}}
\put(40,60){\vector(-1,1){20}}
\thinlines
\put(72,58){$|H\rangle$}
\put(60,85){$|F\rangle$}
\put(35,95){$|V\rangle$}
\put(7,85){$|S\rangle$}
\put(15,15){$(a)$}
\put(130,118){\vector(1,0){56}}
\put(130,118){\vector(1,0){10}}
\put(158,90){\vector(0,1){56}}
\put(158,90){\vector(0,1){10}}
\put(150,110){\line(1,1){16}}
\put(150,110){\framebox(16,16)}
\put(165,130){\scriptsize HV-PBS}
\put(135,75){$(b)$}
\put(130,28){\vector(1,0){56}}
\put(130,28){\vector(1,0){10}}
\put(158,0){\vector(0,1){56}}
\put(158,0){\vector(0,1){10}}
\put(150,20){\line(1,1){16}}
\put(150,20){\framebox(16,16)}
\put(158,28){\circle{16}}
\put(165,40){\scriptsize FS-PBS}
\put(135,-15){$(c)$}
\end{picture}
\end{center}
\caption{$(a)$ Orientations of the HV and
FS polarization bases used throughout the text. The FS basis is
rotated $+45^{o}$ with respect to the HV basis.  $(b)$ and $(c)$ show the
symbols used to represent polarizing beam splitters in the HV and FS
 bases, respectively. } 
\label{fig:polarizations} 
\end{figure}
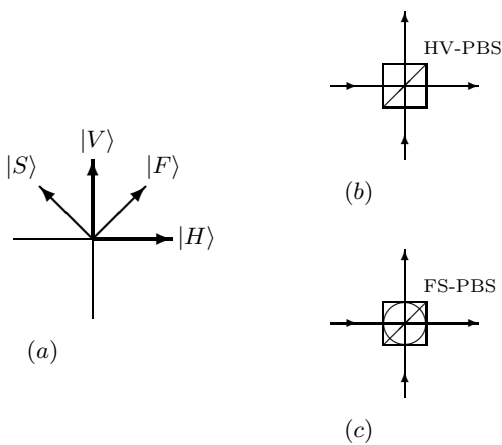

The polarization conventions
that will be used throughout the paper are shown in
Figure \ref{fig:polarizations}.   The photonic qubits $|0\rangle$ and
$|1\rangle$ will be represented by horizontal $|H\rangle$ and 
 vertical
$|V\rangle$ polarizations, respectively, but measurements
will also be
made in the $|F\rangle$ and $|S\rangle$ basis
shown in the figure \cite{fs}. 
Polarizing beam splitters oriented in the HV basis will always transmit
H-polarized photons and reflect V-polarized photons, while
polarizing beam splitters oriented in the FS basis will transmit F-polarized
photons and reflect S-polarized photons.
It will be assumed throughout the paper that the beam splitters and detectors are
ideal and that all the photons in a given optical path are in the same spatial
mode \cite{fearn89,zukowski95,ou97,rarity97}.

\begin{figure}[t]
\vspace*{.5in}
\begin{center}
\begin{picture}(170,210)
\put(10,194){\vector(1,0){80}}
\put(10,194){\vector(1,0){20}}
\put(60,147){\vector(0,1){80}}
\put(60,147){\vector(0,1){15}}
\put(52,186){\line(1,1){16}}
\put(52,186){\framebox(16,16)}
\put(90,191){$\supset$}
\put(96,191){$\sim$}
\put(100,197){$ D_{q}$}
\put(57,227){$\cap$}
\put(59,234){$\wr$}
\put(64,235){$D_{p}$}
\put(75,197){$ q$}
\put(63,158){\mbox{\small 1}}
\put(63,214){{$ p$}}
\put(82,147){\mbox{\small 2}}
\put(82,122){\mbox{\small 3}}
\put(-12,192){{$ |in\rangle_{A}$}}
\put(141,142){{$ |out\rangle_{2}$}}
\put(141,126){{$ |out\rangle_{3}$}}
\put(-13,78){{$ |in\rangle_{B}$}}
\put(-10,207){control}
\put(-10,95){target}
\put(60,137){\circle{20}}
\put(54,134){$|\chi\rangle$}
\put(66,145){\vector(1,0){75}}
\put(66,129){\vector(1,0){75}}
\put(66,145){\vector(1,0){15}}
\put(66,129){\vector(1,0){15}}
\put(10,80){\vector(1,0){80}}
\put(10,80){\vector(1,0){15}}
\put(60,127){\vector(0,-1){80}}
\put(60,127){\vector(0,-1){15}}
\put(52,88){\line(1,-1){16}}
\put(52,72){\framebox(16,16)}
\put(90,77){$\supset$}
\put(96,77){$\sim$}
\put(100,84){$ D_{n}$}
\put(57,41){$\cup$}
\put(59,35){$\wr$}
\put(64,30){$D_{m}$}
\put(75,72){$ n$}
\put(63,60){{$ m$}}
\put(63,111){{$ 4$}}
\end{picture}
\end{center}
\vspace*{-.5in}
\caption{A polarization-encoded  version of the
 Gottesman-Chuang protocol. This scheme relies on a four-photon
 entangled  state, $|\chi\rangle$, to perform a probabilistic
 {\scriptsize CNOT} operation on the two input photonic qubits in
 modes {\small A} and {\small B}. Four polarization-sensitive detectors 
 in the FS basis are
 labelled $D_{m}$, $D_{n}$, $D_{p}$, and $D_{q}$.}
\label{fig:gc} 
\end{figure}
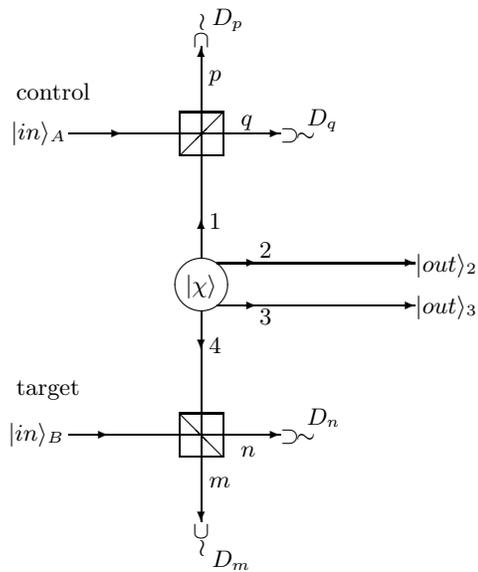

Figure \ref{fig:gc} shows an implementation of a {\scriptsize CNOT} gate using
the GC protocol \cite{gottesman99} with polarization-encoded photonic
qubits and  partial Bell-state measurements.  At the center of the
figure is a source which  emits the  4-photon entangled state $|\chi\rangle$,
whose properties we will now investigate.  The state $|\chi\rangle$ is created
in such a 
way that one photon is emitted into each of the four modes 
labelled {\small 1} through {\small 4}.

The input photon in mode {\small A} is mixed with the photon in mode {\small
1} by a
polarizing beamsplitter in the HV basis with 
 ideal single-photon detectors, $D_{p}$ and
$D_{q}$, in its output ports.  The photons in modes {\small B} and
{\small 4} are mixed in another polarizing beamsplitter in a similar way.  The
remaining two photons of $|\chi\rangle$ in modes {\small 2} and {\small 3}
will  serve as the output qubits.

In a {\scriptsize CNOT} gate the value of the target qubit is to be reversed 
(0 $\leftrightarrow$ 1) if the control qubit has the value 1.
If the input photons in modes {\small A} and {\small B} are considered the
``control'' and ``target'' photons, respectively, then
the desired {\scriptsize CNOT} gate
operation corresponds to the following state transformation:

\begin{eqnarray}
\alpha_{1}|H_{A}\rangle|H_{B}\rangle +
\alpha_{2}|H_{A}\rangle|V_{B}\rangle +
\alpha_{3}|V_{A}\rangle|H_{B}\rangle +
\alpha_{4}|V_{A}\rangle|V_{B}\rangle 
\nonumber
\\
\rightarrow
\alpha_{1}|H_{2}\rangle|H_{3}\rangle +
\alpha_{2}|H_{2}\rangle|V_{3}\rangle +
\alpha_{3}|V_{2}\rangle|V_{3}\rangle +
\alpha_{4}|V_{2}\rangle|H_{3}\rangle 
\nonumber
\\ 
\label{cnot1}
\end{eqnarray}

\noindent
where the $\alpha$ are arbitrary coefficients
($\sum_{i=1}^{4}|\alpha_{i}|^{2}=1$), and, for example, $|H_{A}\rangle$ denotes
a single horizontally polarized photon in mode {\small A}.  From here out, the
kets will be dropped from the notation for simplicity.

In Figure \ref{fig:gc}, the partial Bell measurements 
at the upper and lower
beamsplitters simply consist of only accepting the outputs in modes
{\small 2} and
{\small 3} if one-and-only-one (1AO1) photon is detected in each of the four
detectors.  Combined with the polarization-dependent reflections and
transmissions at the beamsplitters, this condition allows the required properties of the state
$\chi$ to be simply read off from the {\scriptsize CNOT} state transformation in equation
(\ref{cnot1}). 

For example, for the input amplitude $\alpha_{1}H_{A}H_{B} $,
the partial Bell measurements will only succeed if each of the photons in modes
{\small 1} and {\small 4} are also H-polarized (having one of them V-polarized
would result in two photons at one detector and zero at another).
Furthermore, the {\scriptsize CNOT} transformation for this particular input 
amplitude requires the output photons in modes
{\small 2} and
{\small 3} to be H-polarized, so that the state $\chi$ must contain
an amplitude of the form $H_{1}H_{4}H_{2}H_{3}$. 

 The three remaining 
amplitudes can be read off in a similar manner to reveal the required form of
$\chi$:

\begin{equation}
\chi \! =\!
\frac{1}{2} 
\left( 
H_{1}H_{4}H_{2}H_{3} \!+\!
H_{1}V_{4}H_{2}V_{3} \!+\!
V_{1}H_{4}V_{2}V_{3} \!+\!
V_{1}V_{4}V_{2}H_{3} 
\right)
\label{chi}
\end{equation}

\noindent 
It can be seen from this argument that the required polarizations will be
generated in the output modes whenever each of the detectors registers
1AO1 photon.

Although the above argument determines the required form of the state $\chi$,
we must also consider the way in which the output photons are entangled with the
polarizations of the photons in the paths $m$, $n$, $p$, and $q$ leading to the
detectors.  Any such entanglement would result in a more complicated final state than that
shown in Equation (\ref{cnot1}).  Roughly speaking, entanglement of this kind
would provide information regarding the state of the input qubits, which would destroy the
coherence of a quantum computing algorithm.  This situation is analyzed in detail
in the Appendix, where it is shown that this kind of information can be 
``erased'' if each of the four detectors measure the polarization of the photons
in the FS basis.  This is the case because an F-polarized photon is an equal superposition
of H and V polarizations, for example, so that such a measurement provides  no
information regarding the original values of the qubits.  In this particular
example, the use of a quantum erasure technique of this kind is equivalent to a 
$|\phi^{+}\rangle$ Bell-state measurement \cite{michler96}, but that is not the
case for the more general quantum erasure operations described in Section III below.

It is shown in the appendix that the specified output from the detectors will be
obtained $\frac{1}{4}$ of the time, so that this {\scriptsize CNOT} gate
succeeds with a probability of 25\%, provided that the state $\chi$ has been
created with certainty.

\section{{\small CNOT} using Two-Photon Entangled States} 

In this section, we describe
the implementation of several kinds of probabilistic quantum logic gates
using polarizing beam splitters.  We then combine these elementary operations to
give an implementation of a {\scriptsize CNOT} gate that does not rely upon
quantum
teleportation in any obvious way.
 In addition to providing additional insight into the nature of probabilistic
  quantum logic gates, this implementation may have some practical
advantages in certain applications, as will be discussed in the next section.  In
particular, the {\scriptsize CNOT} described here uses a two-photon entangled
state as a resource,  rather than the four-photon
entangled state required for the {\scriptsize CNOT} of the previous section.

\subsection{Quantum Parity Check} 

The first logic operation of interest is the probabilistic quantum parity check
shown in Figure $3$.  The goal of this device is to transfer the value of the
input qubit in mode $2^{\prime}$ to the output in mode $2$, provided that its
value is the same as that of a second input qubit in mode $a$ (even parity). 
The device fails and no output is produced if the two qubits have different values 
(odd parity), and one
of the qubits is destroyed in any event.  The operation of this device does not
measure or determine the values of either input qubit.  The input and output
modes
 have been labeled in this way to facilitate the subsequent construction of a
 {\scriptsize CNOT} gate.
 Pan et al. \cite{pan01a} have independently proposed the use of
polarizing beam splitters to compare the polarization of two photons for
use in entanglement purification, and similar parity checks have been proposed
for other applications
\cite{pan98a,pan01b,bouwmeester01,yamamoto01,zhao01}.

\begin{figure}[b]
\begin{center}
\begin{picture}(200,170)
\put(10,60){\vector(1,0){100}}
\put(10,60){\vector(1,0){20}}
\put(60,10){\vector(0,1){100}}
\put(60,10){\vector(0,1){20}}
\put(52,52){\line(1,1){16}}
\put(52,52){\framebox(16,16)}
\put(52,110){\dashbox{2}(16,16){$ D_{c}$}}
\put(85,51){{$ 2$}}
\put(63,30){{$ a$}}
\put(15,51){{$ 2^{\prime}$}}
\put(63,90){{$ c$}}
\put(-8,58){{$ |in\rangle$}}
\put(112,57){{$ |out\rangle$}}
\put(135,85){\dashbox{2}(55,70)}
\put(144,109){\framebox(12,12)}
\put(150,115){\circle{12}}
\put(144,109){\line(1,1){12}}
\put(150,115){\vector(1,0){15}}
\put(150,90){\vector(0,1){40}}
\put(143,93){$ c$}
\put(147,130){$\cap$}
\put(149,137){$\wr$}
\put(154,142){{$D_{c}^{F}$}}
\put(165,113){$\supset$}
\put(171,113){$\sim$}
\put(174,120){{$ D_{c}^{S}$}}
\end{picture}
\end{center}
\vspace*{-.25in}
\caption{Implementation of a probabilistic quantum parity check of the qubits
in mode $2^{\prime}$
and mode $a$ using a polarizing beam splitter in the HV basis and a
polarization-sensitive detector $D_{c}$ in the FS basis.  The dashed-box insert
shows the details of $D_{c}$, which consists of a polarizing beam splitter in the FS basis
followed by two ordinary single-photon detectors.}
\label{fig:ms1} 
\end{figure}
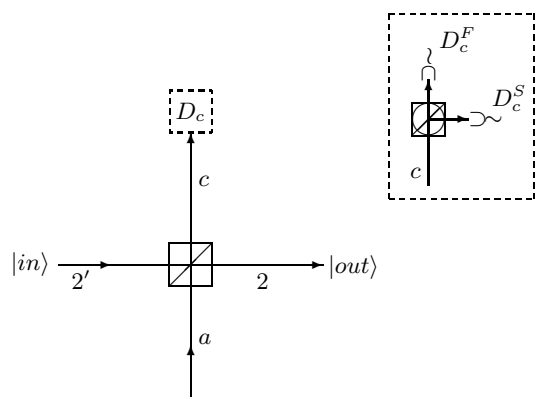 

As illustrated in Figure \ref{fig:ms1}, the quantum parity check can be
implemented by  mixing the photons in a polarizing
beam splitter and accepting the output in mode {\small 2} only for those cases
in which polarization-sensitive detector $D_{c}$ recieves 1AO1 photon. 
This can only occur if the two photons have the same
polarization.
In order to preserve the coherence of an arbitrary input 
superposition
state, however, the polarization-sensitive detector $D_{c}$ consists of a
polarizing beamsplitter oriented in the FS basis and two ordinary single photon
detectors as shown
in the inset of Figure \ref{fig:ms1}.

The operation of the quantum parity check can be understood as follows:
Consider an arbitrary polarization state of the input qubit
$|in\rangle_{2^{\prime}} =\alpha H_{2^{\prime}}+\beta V_{2^{\prime}}$, 
and a single  photon in the
 state $|\phi_{a}\rangle =
\frac{1}{\sqrt{2}}(H_{a}+V_{a})$. The total state,
 $\Psi_{ 2^{\prime}a} \equiv|in\rangle_{2^{\prime}} \otimes
 |\phi_{a}\rangle$,
 is transformed by the polarizing beamsplitter as follows:

\begin{equation}
\Psi_{ 2^{\prime}a}
\longrightarrow \frac{\alpha}{\sqrt{2}}H_{2}H_{c}+
\frac{\beta}{\sqrt{2}}V_{2}V_{c}+
\frac{1}{\sqrt{2}}\psi_{I}
\label{ms11}
\end{equation}

\noindent  where $\psi_{I} \equiv \alpha H_{2}V_{2} +\beta H_{c}V_{c}$ is
composed of amplitudes that will lead to unsuccesful cases in which
$D_{c}$ recieves
two or zero photons.

Using the polarization conventions shown in Figure \ref{fig:polarizations}(a) to write the
mode $c$ amplitudes in the FS basis leads to:

\begin{equation}
\Psi_{ 2^{\prime}a} 
\rightarrow 
\frac{1}{2} \left[ F_{c} (\alpha H_{2} + \beta V_{2}) 
+ S_{c} (-\alpha H_{2} + \beta V_{2}) \right]
+\frac{1}{\sqrt{2}}\psi_{I}
\label{ms12}
\end{equation}

\noindent From the first term inside the square brackets of equation
(\ref{ms12}), we see that
if we passively accept the output only when we recieve 1AO1
photon  in $D_{c}^{F}$ and zero photons in $D_{c}^{S}$, the arbitrary
polarization of the input photon in mode $2^{\prime}$ will be mapped onto
 the photon in
mode {\small 2} with a probability of success equal to  $\frac{1}{4}$.

Furthermore, it can be seen from the second term inside the square brackets of
equation (\ref{ms12}) that we can improve this probability from 
  $\frac{1}{4}$ to  $\frac{1}{2}$ if we
also accept the case when we recieve 1AO1 photon in
$D_{c}^{S}$ and zero photons in $D_{c}^{F}$.  In this case however, we need to
actively
impart an additional phase shift that transforms mode {\small 2} in the
following way: $-\alpha H_{2} + \beta V_{2} \rightarrow \alpha H_{2} + \beta
V_{2}$.

Note that accepting either of these two detection outcomes does not provide any
type of ``which-polarization'' information that would essentially serve to
measure the input.  Imparting the additional $\pi$ phase shift could be done
by rapidly changing the bias voltage on a Pockel's cell in mode 2 upon
detection of a photon in $D_{c}^{S}$.
The quantum erasure technique used here to eliminate any knowledge of the input
qubit is not equivalent to a Bell-state measurement or to quantum teleportation,
since it only involves the detection of a single photon.

\subsection{Destructive-{\bf {\scriptsize CNOT}}} 

\begin{figure}[t]
\vspace*{-.2in}
\begin{center}
\begin{picture}(200,170)
\put(10,100){\vector(1,0){100}}
\put(10,100){\vector(1,0){20}}
\put(60,150){\vector(0,-1){100}}
\put(60,150){\vector(0,-1){20}}
\put(52,108){\line(1,-1){16}}
\put(52,92){\framebox(16,16)}
\put(60,100){\circle{16}}
\put(52,34){\dashbox{2}(16,16){$ D_{d}$}}
\put(93,91){$3$}
\put(63,75){{$ d$}}
\put(15,91){$3^{\prime}$}
\put(63,130){{$ b$}}
\put(-8,98){{$ |in\rangle$}}
\put(114,98){{$ |out\rangle$}}
\put(-15,115){target}
\put(68,140){control}
\put(135,-1){\dashbox{2}(55,70)}
\put(144,29){\framebox(12,12)}
\put(144,41){\line(1,-1){12}}
\put(150,35){\vector(1,0){15}}
\put(150,60){\vector(0,-1){40}}
\put(143,52){$ d$}
\put(147,14){$\cup$}
\put(149,8){$\wr$}
\put(154,3){{$D_{d}^{H}$}}
\put(165,33){$\supset$}
\put(171,33){$\sim$}
\put(174,40){{$ D_{d}^{V}$}}
\end{picture}
\end{center}
\vspace*{-.2in}
\caption{Implementation of a destructive-{\scriptsize CNOT} that performs
a state-flip on the photon in mode $3^{\prime}$ that is 
controlled by the polarization of a single photon in mode $b$. 
Its successful operation
requires the destruction of the control photon.   
The dashed-box inset shows the details of the polarization-sensitive 
detector $D_{d}$.} 
\label{fig:ms2} 
\end{figure}
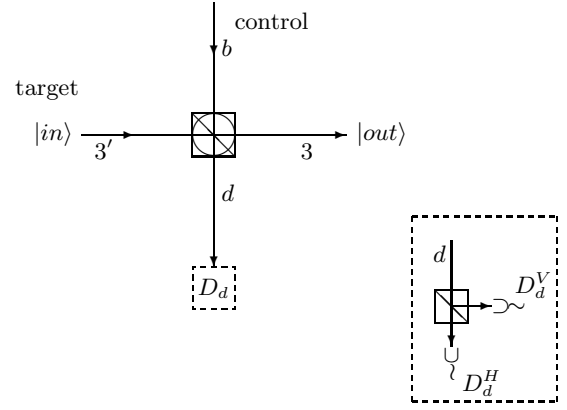

A probabilistic logic device that we refer to as a 
destructive-{\scriptsize CNOT} is shown in Figure
\ref{fig:ms2}.  In this device, the ``target'' input photon in mode 
$3^{\prime}$ 
is mixed with another input photon in mode $b$ at a
polarizing beamsplitter which is oriented in the FS basis.
The goal of this device is to flip the polarization state (eg. H
$\leftrightarrow$ V) of the ``target'' photon if the ``control'' photon 
in mode $b$ is
V-polarized,  and do nothing if it is H-polarized.
This operation is equivalent to a probabilistic {\scriptsize CNOT}
 gate except that the information contained in the control
photon is destroyed in the process.

As in the quantum parity check, the output in mode {\small 3} will only
be accepted if the polarization-sensitive detector $D_{d}$ (see inset box in
Figure \ref{fig:ms2}) receives  1AO1 photon.  The operation of Figure
\ref{fig:ms2} can be understood 
by first considering the
 case of an arbitrary polarization state of the ``target'' photon,
 $|in\rangle_{3^{\prime}} =\alpha H_{3^{\prime}}+\beta V_{3^{\prime}}$, and a
 single V-polarized photon 
in mode $b$, $|\phi_{b}\rangle = V_{b}$.
 
 Writing these states in the FS basis, the total state 
 $\Psi_{3^{\prime}b} \equiv |in\rangle_{3^{\prime}} \otimes |\phi_{b}\rangle $
 is given by:

\begin{equation}
\Psi_{3^{\prime}b}
=
\left[ \frac{\alpha}{\sqrt{2}}(F_{3^{\prime}}-S_{3^{\prime}}) +
\frac{\beta}{\sqrt{2}}(F_{3^{\prime}}+S_{3^{\prime}}) 
\right]
\otimes
\frac{1}{\sqrt{2}}(F_{b}+S_{b})
\label{ms21}
\end{equation}

\noindent The polarizing beamsplitter transforms this into:

\begin{equation}
\Psi_{3^{\prime}b}
\rightarrow  
\frac{1}{2} 
\left[ 
\alpha(F_{d}F_{3} - S_{d}S_{3}) +
\beta(F_{d}F_{3} + S_{d}S_{3})
\right]
+\frac{1}{\sqrt{2}}\psi_{II}
\label{ms22}
\end{equation}

\noindent where
 $\psi_{II} \equiv
\frac{\alpha}{\sqrt{2}}(F_{3}S_{3} - F_{d}S_{d}) +
\frac{\beta}{\sqrt{2}}(F_{3}S_{3} + S_{d}F_{d})$
includes the amplitudes that would lead to the unsuccessful cases in
which $D_{d}$ does not recieve 1AO1 photon.  Rewriting the amplitudes in mode
$d$ back in the HV basis leads to:

\begin{equation}
\Psi_{3^{\prime}b}
\rightarrow  
\frac{1}{2} 
\left[ 
H_{d}(\alpha V_{3} +\beta H_{3}) +
V_{d}(\alpha H_{3} +\beta V_{3}) 
\right]
+\frac{1}{\sqrt{2}}\psi_{II}
\label{ms23}
\end{equation}

If we accept only outcomes in which detector $D_{d}^{H}$ recieves 1A01
photon and $D_{d}^{V}$ recieves zero photons, the output collapses to the state
$\alpha V_{3} +\beta H_{3}$.  This corresponds to a ``flip''  of the input
state $\alpha H_{3^{prime}} +\beta V_{3^{\prime}}$ and occurs with a
probability of $\frac{1}{4}$.  

From equation (\ref{ms23}), we see that this probability can be increased to
$\frac{1}{2}$ if we also accept the outcomes in which $D_{d}^{V}$ recieves 1A01
photon and $D_{d}^{H}$ recieves zero photons. For this outcome, however, we
need to apply  classically controlled single-qubit operations that accomplish the
transformation $\alpha H_{3} +\beta V_{3} \rightarrow \alpha V_{3} +\beta
H_{3}$.  This could be done, for example, by first rotating the polarization of
the photon in mode {\small 3} by $90^{o}$, and then imparting the same type of
polarization dependent $\pi$-phase shift that was used in quantum parity check
device.

To complete the description of the destructive-{\scriptsize CNOT}
operation,  we now consider the case
where we have the same arbitrary state of the input ``target'' photon, 
$|in\rangle_{3^{\prime}} =\alpha H_{3^{\prime}}+\beta V_{3^{\prime}}$, but a
 single H-polarized photon in mode $b$, $|\phi_{b}\rangle = H_{b}$.  Following
 the  steps that lead to equation (\ref{ms23}), it can be shown that:

\begin{equation}
\Psi^{\prime}_{3^{\prime}b}
\rightarrow 
\frac{1}{2} 
\left[ 
H_{d}(\alpha H_{3} +\beta V_{3}) +
V_{d}(\alpha V_{3} +\beta H_{3}) 
\right]
+\frac{1}{\sqrt{2}}\psi_{III}
\label{ms24}
\end{equation}

\noindent where  
$\psi_{III} \equiv
\frac{\alpha}{\sqrt{2}}(F_{3}S_{3} + F_{d}S_{d}) +
\frac{\beta}{\sqrt{2}}(F_{3}S_{3} - S_{d}F_{d})$
again includes the amplitudes that lead to unsuccessful detection.
In comparison with equation (\ref{ms23}), we see that using the same detection
scheme and single-qubit operations for the case of an H-polarized control
photon will leave the state of the target photon unchanged.

From equations (\ref{ms23}) and (\ref{ms24}) it is clear that the 
destructive-{\scriptsize CNOT} performs a
polarization
state-flip transformation of the input photon in mode $3^{\prime}$ that 
is controlled by  the polarization of the control photon in mode $b$.  
However, this transformation is only realized by a post-selection
process that destroys the control photon.

\subsection{Quantum Encoder} 

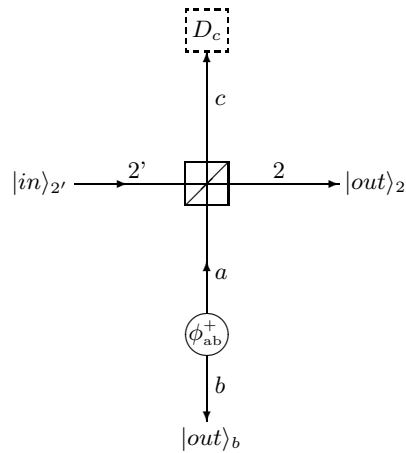
\begin{figure}[t]
\begin{center}
\begin{picture}(140,170)
\put(10,94){\vector(1,0){100}}
\put(10,94){\vector(1,0){20}}
\put(60,45){\vector(0,1){99}}
\put(60,45){\vector(0,1){20}}
\put(52,86){\line(1,1){16}}
\put(52,86){\framebox(16,16)}
\put(52,144){\dashbox{2}(16,16){$ D_{c}$}}
\put(85,96){\mbox{\small 2}}
\put(63,58){{$ a$}}
\put(63,15){{$ b$}}
\put(30,96){\mbox{\small 2'}}
\put(63,124){{$ c$}}
\put(-14,92){{$ |in\rangle_{2^{\prime}}$}}
\put(112,92){{$ |out\rangle_{2}$}}
\put(50,-5){{$ |out\rangle_{b}$}}
\put(60,37){\circle{16}}
\put(53,35){$\phi_{\mbox{\tiny ab}}^{\mbox{\tiny +}}$}
\put(60,29){\vector(0,-1){25}}
\end{picture}
\end{center}
\caption{A quantum encoder circuit that is identical to the quantum
parity-check of Figure \ref{fig:ms1} except the single photon 
in mode $a$ is
now part of a two-photon Bell-state 
$\phi^{+}_{ab}=\frac{1}{\sqrt{2}}(H_{a}H_{b} + V_{a}V_{b})$.}
\label{fig:ms3} 
\end{figure}

Although the {\scriptsize CNOT} described above destroys the control photon,
such a device would still be useful 
if we could copy the value of the control qubit before the
{\scriptsize CNOT} operation, so that all of the 
information would still be available after the
operation was completed.  With that in mind, we now describe the operation of a
probabilistic quantum encoder as shown in Figure \ref{fig:ms3}.  The intended
function of this device is to copy (encode) the value of the input qubit in mode 
$2^{\prime}$  into
both of the output modes $2$ and $b$.  An operation of this kind is not
equivalent to quantum cloning \cite{wooters82} and is conventionally referred
to as a quantum encoder \cite{chuangneilsen}.
  Once again, the modes have been labeled in such
a way as to facilitate the subsequent construction of a {\scriptsize CNOT}.

A quantum encoder can be implemented by combining the quantum parity check
described above with a two-photon entangled state of the form 
$\phi^{+}_{ab}=\frac{1}{\sqrt{2}}(H_{a}H_{b} + V_{a}V_{b})$, where one
member of the two-photon entangled state provides the input to mode $a$ of the
quantum parity check.  This is illustrated in Figure \ref{fig:ms3}, where we
use the same detection scheme and single
qubit operations 
used in the quantum parity check.

Recall that in the quantum parity check, a successful detection event
post-selected outcomes which effectively realized the transformation 
$\alpha H_{2^{\prime}}+\beta V_{2^{\prime}}
\rightarrow 
\alpha H_{2}+\beta V_{2}$, provided that the values of the qubits in modes $a$
and $2^{\prime}$ were the same.
  Based on the same arguments used to derive equation (\ref{ms12}), it can be
shown that successful detection in Figure \ref{fig:ms3} post-selects the
following transformation of the same arbitrarily 
polarized input state:

\begin{equation}
\alpha H_{2^{\prime}}+\beta V_{2^{\prime}}
\rightarrow
\alpha H_{2}H_{b}+\beta V_{2}V_{b}
\label{ms31}
\end{equation}

\noindent As in the quantum parity check, the operation of the quantum encoder
succeeds with a probability of $\frac{1}{2}$.

It should be apparent from equation (\ref{ms31}) that the operation of a
quantum encoder is not equivalent to quantum cloning \cite{wooters82}, even
though the  value of the input qubit is
copied onto two output qubits.  The value of the input and output qubits is not
determined or measured during the operation of this device.

\subsection{Non-Destructive {\bf {\scriptsize CNOT}}} 

\begin{figure}[t]
\begin{center}
\begin{picture}(210,270)
\put(60,194){\vector(1,0){100}}
\put(60,194){\vector(1,0){20}}
\put(110,145){\vector(0,1){99}}
\put(110,145){\vector(0,1){20}}
\put(102,186){\line(1,1){16}}
\put(102,186){\framebox(16,16)}
\put(102,244){\dashbox{2}(16,16){$ D_{c}$}}
\put(135,196){\mbox{\small 2}}
\put(113,158){{$ a$}}
\put(80,196){\mbox{\small 2'}}
\put(113,224){{$ c$}}
\put(36,192){{$ |in\rangle_{2^{\prime}}$}}
\put(162,192){{$ |out\rangle_{2}$}}
\put(15,205){control}
\put(110,137){\circle{16}}
\put(103,135){$\phi_{\mbox{\tiny ab}}^{\mbox{\tiny +}}$}
\put(60,80){\vector(1,0){100}}
\put(60,80){\vector(1,0){20}}
\put(110,129){\vector(0,-1){99}}
\put(110,129){\vector(0,-1){20}}
\put(102,88){\line(1,-1){16}}
\put(102,72){\framebox(16,16)}
\put(110,80){\circle{16}}
\put(102,14){\dashbox{2}(16,16){$ D_{d}$}}
\put(135,82){\mbox{\small 3}}
\put(113,55){{$ d$}}
\put(80,82){\mbox{\small 3'}}
\put(113,112){{$ b$}}
\put(36,78){{$ |in\rangle_{3^{\prime}}$}}
\put(164,78){{$ |out\rangle_{3}$}}
\put(15,90){target}
\put(130,112){\scriptsize (Encoder)}
\put(130,37){{\scriptsize (Destructive-CNOT)}}
\put(130,162){\scriptsize (Parity-Check)}
\end{picture}
\end{center}
\vspace*{-.25in}
\caption{Construction of a probabilistic {\scriptsize CNOT} gate by combining
a quantum encoder with a destructive-{\scriptsize CNOT}. 
The mode labels have been preserved to facilitate
comparisons with Figures \protect\ref{fig:ms2} and \protect\ref{fig:ms3}.}
\label{fig:cnot} 
\end{figure}
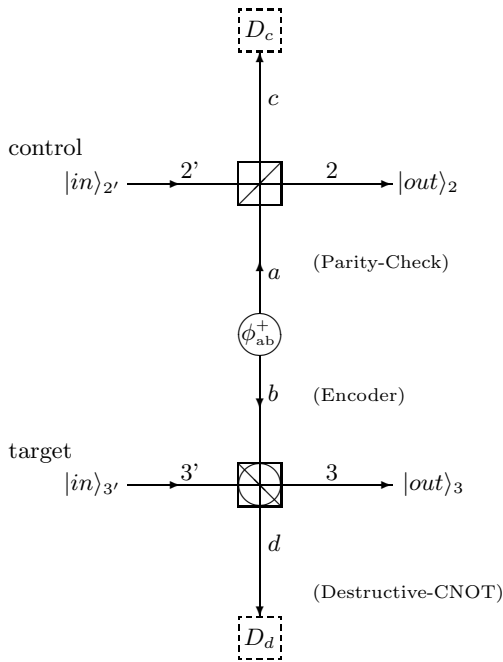

The elementary logic operations described above can now be combined to implement
a non-destructive {\scriptsize CNOT} gate as illustrated in Figure
\ref{fig:cnot}. The quantum encoder of Figure \ref{fig:ms3} is used to copy the
value of the control qubit of mode $2^{\prime}$ directly into the output in mode $2$ as
well as into the input mode $b$ of the destructive-{\scriptsize CNOT} gate of 
Figure \ref{fig:ms2}.   The output in mode $3$ will then contain the result of
the destructive-{\scriptsize CNOT} operation while the value of the control
qubit  is preserved in the other output
mode.  All of the mode labels have been preserved to facilitate comparisons of
Figures \ref{fig:ms2}, \ref{fig:ms3}, and \ref{fig:cnot}.

Since the quantum parity check and destructive-{\scriptsize CNOT} operations
both  succeed with
a probability of $\frac{1}{2}$, it follows that the {\scriptsize CNOT}
 gate of Figure \ref{fig:cnot} will
succeed with an overall probability of $\frac{1}{4}$.  The operation of the
gate can be explicitly verified by 
considering an arbitrary input state of the form:

\begin{equation}
\Psi_{2^{\prime}3^{\prime}} = 
\alpha_{1}H_{2^{\prime}}H_{3^{\prime}} +
\alpha_{2}H_{2^{\prime}}V_{3^{\prime}} +
\alpha_{3}V_{2^{\prime}}H_{3^{\prime}} +
\alpha_{4}V_{2^{\prime}}V_{3^{\prime}},
\label{psiin}
\end{equation}

\noindent and calculating the evolution of the total state $\Psi_{t} \equiv 
\Psi_{2^{\prime}3^{\prime}}\otimes \phi^{+}_{ab}$.  Based on the same types of 
arguments used in the descriptions of the earlier logic devices, it can be
shown that:

\begin{eqnarray}
\Psi_{t}\!&\rightarrow \!
\frac{1}{4}&\!
\{
F_{c}H_{d} \left[
+\alpha_{1}H_{2}H_{3} \!+\!
\alpha_{2}H_{2}V_{3} \!+\!
\alpha_{3}V_{2}V_{3} \!+\!
\alpha_{4}V_{2}H_{3} \right]
\nonumber\\
&&+S_{c}H_{d} \left[
-\alpha_{1}H_{2}H_{3} 
-\alpha_{2}H_{2}V_{3} \!+\!
\alpha_{3}V_{2}V_{3} \!+\!
\alpha_{4}V_{2}H_{3} \right]
\nonumber\\
&&+F_{c}V_{d} \left[
+\alpha_{1}H_{2}V_{3} \!+\!
\alpha_{2}H_{2}H_{3} \!+\!
\alpha_{3}V_{2}H_{3} \!+\!
\alpha_{4}V_{2}V_{3} \right]
\nonumber\\ 
&&+S_{c}V_{d} \left[
-\alpha_{1}H_{2}V_{3} 
-\alpha_{2}H_{2}H_{3} \!+\!
\alpha_{3}V_{2}H_{3} \!+\!
\alpha_{4}V_{2}V_{3} \right]
\}
\nonumber \\
&&+\frac{\sqrt{3}}{2}\psi_{IV}
\label{evolution}
\end{eqnarray}

\noindent where $\psi_{IV}$ is a normalized combination of all of the amplitudes
that would not lead to 1AO1 photon in each of the polarization-sensitive
detectors, 
$D_{c}$ and $D_{d}$. The amplitudes of interest are grouped together 
as the four main terms inside
the curly brackets of equation (\ref{evolution}).

From the first of these four terms, we see that if we passively
accept only those cases in which we recieve 1AO1 photon in $D_{c}^{F}$ and 1AO1
photon in $D_{d}^{H}$ (eg. the passive success conditions of the quantum parity
check device and the destructive-{\scriptsize CNOT}, respectively), 
then we accomplish the desired {\scriptsize CNOT} transformation
on the input state of equation (\ref{psiin}) with a success probability of
$\frac{1}{16}$.
However, if we also accept the three other terms inside
the curly brackets of equation (\ref{evolution})
 and implement the appropriate combinations of the
classically controlled single-qubit operations used in
the quantum parity
check and the destructive-{\scriptsize CNOT},
 then
the  four main terms 
all combine and increase this probability from $\frac{1}{16}$ to
$\frac{1}{4}$.

For example, in the second term inside
the curly brackets of equation (\ref{evolution}), the detection of a single
S-polarized photon in detector $D_{c}$ triggers the 
polarization-dependent $\pi$-phase shift on mode {\small 2}
($H_{2}\rightarrow -H_{2}$).  In the third term, the detection of a single
V-polarized photon in $D_{d}$ requires the state-flip on the photon in  mode {\small 3}
(eg. $H_{3}\leftrightarrow V_{3}$).  Finally, we see that the
fourth term inside
the curly brackets of equation (\ref{evolution})
 requires both of these single-qubit operations.

The net result is that the desired {\scriptsize CNOT} operation can be performed with a
probability of $\frac{1}{4}$ without measuring or determining the values of
the input qubits.  The operation is known to be successful with certainty when the
specified combination of detectors each registers a single photon, as is also
the case for previous implementations \cite{knill01,koashi01}.

\section{Summary and Conclusions} 

One of the main results of this paper is the probabilistic {\scriptsize CNOT}
 implementation
shown in Figure \ref{fig:cnot}.  Although our work was inspired by References 
\cite{knill01} and \cite{koashi01}, our {\scriptsize CNOT} 
in Figure \ref{fig:cnot} differs from
earlier implementations in several respects.  Our implementation uses
 only polarizing beam splitters with total reflection or
transmission, which simplifies the analysis.  We showed how a {\scriptsize CNOT}
 gate could be constructed from more elementary
operations, such as a quantum parity check, which also clarifies the operation of
the resulting {\scriptsize CNOT} gate.  Our approach is not directly related to quantum
teleportation and depends, instead, on a quantum erasure technique.  Finally, our
implementation provides a higher probability of success for a given number of
ancilla than is the case for the earlier implementations, and it uses half as
many detectors as do several other implementations.  We now discuss some of these
features in more detail.

From a basic physics perspective, one conclusion from
our work is that quantum erasure can play an essential role in the
implementation of a probabilistic {\scriptsize CNOT} gate.  The implementation of the
elementary logic operations described above were all based on the mixing of two
input photons on a single beam splitter, followed by a single
polarization-sensitive 
detector.  Quantum erasure techniques were used to eliminate any 
information regarding the values of the two inputs.  In contrast, quantum
teleportation uses the output of two single photon detectors to recreate the
state of one of the original photons, as was discussed in connection with the
Gottesman-Chuang protocol in Section II.  The fundamental feature of quantum
logic operations as opposed to classical logic operations is that the input
qubits must remain uncertain, which can be accompished using 
quantum erasure techniques.

The fundamental role of quantum erasure can be further understood by comparing
our implementation in Figure \ref{fig:cnot} with that of Koashi, Yamamoto, and
Imoto \cite{koashi01}.
  Without going into any detail, our Figure \ref{fig:cnot} bears a
(superficial) resemblance to the inner portion of their implementation,
in which they used four ancilla photons to generate a state analogous to the
four-photon entangled state $\chi$.  They followed this by
quantum teleportation of the two input qubits using two Bell-state measurements, which
constituted the remainder of their implementation.  Our implementation is roughly
equivalent to eliminating their teleportation and Bell state measurements while
replacing two of their ancilla with the actual input qubits.  The net result is
that our approach achieves a {\scriptsize CNOT} with a probability of 
$\frac{1}{4}$ using only two ancilla, while the basic KYI implementation 
\cite{koashi01} requires four ancilla and succeeds with a probability 
of $\frac{1}{16}$. Our approach also requires half as many detectors.  
This result can be interpreted as a more
efficient use of quantum erasure without quantum teleportation.

One of the original implementations suggested by Knill, LaFlamme, and Milburn
\cite{knill01}
also involves the generation of a four-mode entangled state analogous to
$\chi$  followed by
partial Bell-state measurements similar to those of Section II.
A {\scriptsize CNOT} gate could also be implemented without teleportation using
their nonlinear phase shift, which would also acheive a
probability of success of $\frac{1}{16}$ using two ancilla, whereas our
implementation achieves a probability of success of $\frac{1}{4}$ using two ancilla.  
As a practical
matter, however, it should be noted that our implementation requires the two
ancilla to be entangled whereas the KLM approach does not.

Although the {\scriptsize CNOT} scheme of Figure \ref{fig:cnot} requires only
two ancilla in contrast to the four-photon entangled state $\chi$ required for
the  {\scriptsize CNOT} scheme of Figure \ref{fig:gc}, the latter
approach does offer some computational advantages.  For example, the reliance
on four ancilla photons in the original KYI proposal \cite{koashi01} allows
them overcome certain
problems associated with imperfect sources and photon loss.  Furthermore, KLM have 
described
 a remarkable process in which the success probability of the
teleportations required in a GC-like protocol can be increased arbitrarily
close to one using more complex linear optics techniques \cite{knill01}.  In
this way, they show that any GC-like protocol could, in principle, be reduced
to the problem of preparing a special 4-qubit state analagous to $\chi$.  The
beauty of the KLM proposal is that this allows the state $\chi$ to be prepared
probabilistically.  Upon successful generation of $\chi$, a GC-like
protocol can then be carried out on the qubits of interest.

\begin{figure}[t]
\vspace*{-.1in}
\begin{center}
\begin{picture}(100,140)
\put(10,50){\circle{16}}
\put(5,48){$ \phi^{+}$}
\put(10,42){\vector(0,-1){40}}
\put(18,50){\vector(1,0){70}}
\put(18,50){\vector(1,0){10}}
\put(11,25){\small 4}
\put(28,42){\small $3^{\prime}$}
\put(78,42){\small 3}
\put(10,80){\circle{16}}
\put(5,78){$ \phi^{+}$}
\put(10,88){\vector(0,1){40}}
\put(18,80){\vector(1,0){70}}
\put(18,80){\vector(1,0){10}}
\put(11,100){\small 1}
\put(28,82){\small $2^{\prime}$}
\put(78,82){\small 2}
\put(50,40){\dashbox{2}(20,50)}
\put(48,92){\scriptsize C-NOT}
\put(60,80){\circle*{5}}
\put(60,80){\line(0,-1){34}}
\put(60,50){\circle{8}}
\end{picture}
\end{center}
\vspace*{-.25in}
\caption{GC method \protect\cite{gottesman99} for creation of the 4-photon
entangled state $\chi$ from two 2-photon entangled states. Here
$\phi^{+}_{ij^{\prime}}=\frac{1}{\sqrt{2}}(H_{i}H_{j^{\prime}}+V
_{i}V_{j^{\prime}})$, where $i,j = 1,2 $ or $4,3$. The {\scriptsize CNOT}
between modes $2^{\prime}$ and $3^{\prime}$ converts the product 
$\phi^{+}_{12^{\prime}}\otimes \phi^{+}_{43^{\prime}}$ into the 4-photon state
$\chi$.}
\label{fig:chi}  
\end{figure}
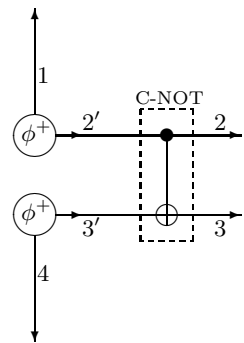

Within this context, it is interesting to note that the probabilistic
{\scriptsize CNOT} described in Figure \ref{fig:cnot} could itself be used to
 generate the state
$\chi$ by a method suggested in reference \cite{gottesman99} (see Figure
\ref{fig:chi}).  Given a reliable source of triggered two-photon Bell-states, 
it would require, on average, only
four attempts to generate the four-photon entangled state $\chi$.
  In this sense, the {\scriptsize CNOT} scheme of Figure
\ref{fig:cnot} could offer a speed-up in the basic state preparation step
for subsequent use in the original KLM proposal
\cite{knill01}.

Experimental realizations of the quantum parity check,
quantum encoder, and destructive-{\scriptsize CNOT} devices used to build the 
 {\scriptsize CNOT} scheme of Figure
\ref{fig:cnot}  will require reliable sources of single photons to serve as the
qubits of interest,
 as well as reliable sources of triggered two-photon entangled states to serve
 as the ancilla.
Several promising approaches towards 
generating single photons on demand are currently
being investigated (see, for example
\cite{kimble77,ou86,demartini96,kim99a,michler00,kurtsiefer00,brouri00,lounis00,bratke01}).
A source of triggered two-photon Bell-states
could, for example, be produced using
  entanglement swapping between two parametric
down-conversion photon pairs
\cite{zukowski93,pan98b}, or more directly using recently discovered
quantum-dot techniques \cite{benson00}.

In summary, 
we have described the operation of several photonic 
quantum logic devices of an elementary nature, including a quantum parity
check, a quantum encoder, and a destructive-{\scriptsize CNOT}.
These devices can be combined to perform a probabilistic
{\scriptsize CNOT} operation on two arbitrary
polarization-encoded photonic qubits.
These quantum logic operations rely on linear optical elements in the form of
polarizing beam splitters, additional photons (ancilla), and a post-selection
process based on the outcome of single photon detectors.  Our results provide
additional insight into the nature of probabilistic logic operations using linear
elements and they suggest that quantum erasure techniques can play a 
fundamental role in these devices.

\vspace*{.15in}
This work was supported by the Office of Naval Research and by internal IR\&D
funds.

\section*{Appendix} 

In Section II, we introduced the basic idea behind a limited
version of the
the GC protocol \cite{gottesman99} that relied on polarization-encoded qubits
and partial Bell-measurements. In this case, the partial Bell-measurement proceedure 
was
simply to accept the output in modes {\small 2} and {\small 3} if and only if
each of the four detectors received 1AO1 photon.  
That approach
allowed the required properties of the special
 4-photon entangled state $\chi$ to be understood in a
straightforwarad manner.

In this Appendix, we provide a more detailed analysis that includes the
entanglement between the output photons and the polarizations of the photons in
the modes $m$, $n$, $p$, and $q$ leading to the four detectors.  It will be
found that this entanglement would provide information regarding the values of the input
qubits unless that information is ``erased'' using polarization sensitive
detectors in the FS basis.  Furthermore,
additional single-qubit
 operations on the output modes are required
  to achieve the highest possible probability of successful operation.

The initial state of the system is given by the product of the
arbitrary input state, $\psi_{in}$, defined by the left-hand side of equation
(\ref{cnot1}), and the state $\chi$ whose properties are given in
equation (\ref{chi}):

\begin{eqnarray}
\Psi_{t} \!& & \!\!\equiv  \! (
\alpha_{1}H_{A}H_{B}\!+\!
\alpha_{2}H_{A}V_{B}\!+\!
\alpha_{3}V_{A}H_{B}\!+\!
\alpha_{4}V_{A}V_{B})  \otimes
\nonumber\\ &&
\frac{1}{2} 
\left( 
H_{1}H_{4}H_{2}H_{3} \! + \!
H_{1}V_{4}H_{2}V_{3} \! + \!
V_{1}H_{4}V_{2}V_{3} \! + \!
V_{1}V_{4}V_{2}H_{3}  \right)
\nonumber\\&&
\label{ap1}
\end{eqnarray}

\noindent 
Expanding equation (\ref{ap1}) and carrying out the polarizing beamsplitter 
transformations on the
amplitudes in modes {\small A}, {\small B}, {\small 1}, and {\small 4},
results in a total of 16 amplitudes.  Of these 16 amplitudes, only four
correspond to a ``successful'' outcome in which there is
1AO1 photon in each of the four detectors; the other 12 amplitudes correspond
to ``unsuccessful'' outcomes in which one or two detectors recieves two
or zero photons.

It is straightforward to show that:

\begin{eqnarray}
\Psi_{t}  = & \frac{1}{2} & \left\{
\alpha_{1} H_{2}H_{3}(H_{p}H_{q}H_{n}H_{m}) +
\alpha_{2} H_{2}V_{3}(H_{p}H_{q}V_{n}V_{m}) \right.
\nonumber\\ &&
\left.
+\alpha_{3} V_{2}V_{3}(V_{p}V_{q}H_{n}H_{m}) +
\alpha_{4} V_{2}H_{3}(V_{p}V_{q}V_{n}V_{m}) \right\}
\nonumber\\&&
+\frac{\sqrt{3}}{2}\psi_{V}
\label{ap2}
\end{eqnarray}

\noindent where the four ``successful'' amplitudes are explicitly shown, and
$\psi_{V}$ is a normalized combination of the 12 ``unsuccessful''
amplitudes.

The potential difficulty due to entanglement is
apparent from the terms inside the curly brackets of equation (\ref{ap2}).  In
principle, we are able to distinguish the polarizations of the two photons in
the output modes {\small 2} and {\small 3} based on the combinations of
the polarizations of the four photons in the detection modes.  This information
essentially serves to measure the output, thereby spoiling the {\scriptsize
CNOT} transformation of the arbitrary input state. 

In order to ``erase'' this information \cite{scully82,kwiat92,herzog95} and
preserve the coherence of the output amplitudes, each of the four polarization-sensitive
detectors will consist of a
polarizing  beamsplitter in
the FS basis and two ordinary single-photon detectors.  This type of
polarization-sensitive detector is shown in
the inset of Figure \ref{fig:ms1}.  

We therefore
 write each  of the detection mode amplitudes in equation (\ref{ap2}) in the FS basis.
This leads to a total of 16 possible detection outcomes that correspond to the
``success''  condition of having 1A01 photon in each of the detector packages.  It is
 convenient to group the amplitudes of the expansion of equation (\ref{ap2}) in
 terms  of these 16 detection outcomes.  The details are straightforward, but 
only  the  first  and  last  two terms  are shown here  to save space:
 
\begin{eqnarray}
\lefteqn{\Psi_{t}  = }
\nonumber\\
&&\frac{1}{8} \left\{ \right.
F_{p}F_{q}F_{n}F_{m} 
[
+\alpha_{1}H_{2}H_{3}\!+\!
\alpha_{2}H_{2}V_{3}\!+\!
\alpha_{3}V_{2}V_{3}\!+\!
\alpha_{4}V_{2}H_{3}
] 
\nonumber\\&&
+F_{p}F_{q}F_{n}S_{m} 
[
-\alpha_{1}H_{2}H_{3}\!+\!
\alpha_{2}H_{2}V_{3}\!-\!
\alpha_{3}V_{2}V_{3}\!+\!
\alpha_{4}V_{2}H_{3}
]
\nonumber\\&&
\: \vdots 
\nonumber\\&&
\: \vdots
\nonumber\\&&
+S_{p}S_{q}S_{n}F_{m} 
[
-\alpha_{1}H_{2}H_{3}\!+\!
\alpha_{2}H_{2}V_{3}\!-\!
\alpha_{3}V_{2}V_{3}\!+\!
\alpha_{4}V_{2}H_{3}
]
\nonumber\\ &&\left.
+S_{p}S_{q}S_{n}S_{m} 
[
+\alpha_{1}H_{2}H_{3}\!+\!
\alpha_{2}H_{2}V_{3}\!+\!
\alpha_{3}V_{2}V_{3}\!+\!
\alpha_{4}V_{2}H_{3}
] \right\} 
\nonumber\\&&
+\frac{\sqrt{3}}{2}\psi_{V}
\label{ap3}
\end{eqnarray}

From the first term inside the curly brackets, we see that passively accepting
only the detection outcome in which each of the four detectors receives
1AO1 $F$-polarized photon projects the desired {\scriptsize CNOT}
transformation of the arbitrary input onto the ouptut modes 
{\small 2} and {\small 3} with a
success probability of $\frac{1}{64}$.  However, employing the same techniques
described in the text for the quantum parity check and
destructive-{\scriptsize CNOT},
 we can
increase this probability  by accepting the other 15 ``successful''
 detection outcomes and performing classically
 controlled single-qubit operations to
correct for the various minus-signs on the amplitudes in the output modes. In
this way all 16 terms contribute to the same outcome and the probability of
success goes
from $\frac{1}{64}$ to $\frac{1}{4}$.

One way to correct the minus-signs in equation (\ref{ap3}) is to
use a protocol in which detection of an $S$-polarized photon in any of the
detectors activates a specific combination of polarization-dependent
$\pi$-phase shifts on the output modes.  For example, the following protocol will achieve
the desired corrections:

\begin{eqnarray}
S_{p} \: or \: S_{q} & \Rightarrow & (H_{2}\rightarrow -H_{2})
\nonumber\\
S_{m} \: or \: S_{n} & \Rightarrow & (H_{2}\rightarrow -H_{2}) \: \: and \: \:
 (V_{3}\rightarrow
-V_{3})
\label{ap4}
\end{eqnarray}

These phase shifts are to be applied independently for each detector outcome. 
For example, if $S_{p}$ and  $S_{q}$ outcomes are both obtained, then the sign
reversal is to be applied twice, which has no net effect in that case.




\begin{thebibliography}{50}
\bibitem{fs}  These axes have been designated F and S 
in analogy with the fast and slow axes of birefringent crystals. In this
paper, however, the HV and FS bases are equivalent.
\end{thebibliography}
\end{document}